**Long-Lived Spin-Polarized Intermolecular Exciplex States in Thermally Activated Delayed Fluorescence-Based Organic Light-Emitting Diodes**

Sebastian Weissenseel, Andreas Gottscholl, Rebecca Bönnighausen,
Vladimir Dyakonov, Andreas Sperlich*
* sperlich@physik.uni-wuerzburg.de

Experimental Physics 6 and Würzburg-Dresden Cluster of Excellence ct.qmat, Julius Maximilian University of Würzburg, 97074 Würzburg, Germany

**Abstract**
Spin-spin interactions in organic light-emitting diodes (OLEDs) based on thermally activated delayed fluorescence (TADF) are pivotal because radiative recombination is largely determined by triplet-to-singlet conversion, also called reverse intersystem crossing (RISC). To explore the underlying process, we apply a spin-resonance spectral hole-burning technique to probe electroluminescence. We find that the triplet exciplex states in OLEDs are highly spin-polarized and show that these states can be decoupled from the heterogeneous nuclear environment as a source of spin dephasing and can even be coherently manipulated on a spin-spin relaxation time scale $T_2^*$ of 30 ns. Crucially, we obtain the characteristic triplet exciplex spin-lattice relaxation time $T_1$ in the range of 50 µs, which far exceeds the RISC time. We conclude that slow spin relaxation rather than RISC is an efficiency-limiting step for intermolecular donor:acceptor systems. Finding TADF emitters with faster spin relaxation will benefit this type of TADF OLEDs.

**Introduction**
The technological development of organic light-emitting diodes (OLEDs) has undergone remarkable progress, and the market share of OLED displays in television and smart device applications is growing steadily. Initially started with fluorescence-based emitters with a maximum internal quantum efficiency (IQE) of 25%, phosphorescent molecular emitters with enhanced spin-orbit coupling (SOC) enabled much higher device efficiencies through singlet-to-triplet intersystem crossing (ISC) (*1–4*). To achieve 100% IQE with molecular systems that do not contain heavy elements, the concept of thermally activated delayed fluorescence (TADF), originally known as E-type delayed fluorescence (*5–8*) has been successfully used in devices and substantially affected OLED development over the past decade (*9–11*). In the TADF or E-type delayed fluorescence process, the first excited singlet state is populated by a thermally activated transition from the first excited triplet state. The mechanism for harvesting non-radiative triplet states is described as reverse ISC (RISC) (*12–16*). Upon injection of electrons and holes from the contacts into the active layer, which consists of donor and acceptor molecules, intermolecular exciplex states are formed. In general, such Coulomb-bound quasiparticles can be in either the singlet ($S$ = 0) or triplet ($S$ = 1) state, where the latter should, in principle, be accurately described by the spin Hamiltonian and can also be selectively controlled, e.g., by an external magnetic field or electron paramagnetic resonance (EPR), to track the population transfer to the singlet manifold. Of the two possible spin orientations of the exciplex state, EPR can selectively influence (flip) only triplet states, while the effect of the spin-flip is observed in a change of a radiative singlet exciplex state recombination. The missing puzzle piece is the connection between these two spin states, as RISC is a first-order spin-forbidden process (*17–20*). However, it can be thermally activated, i.e., as a result of electron-phonon interactions or through the admixture of other spin states, which, for example, enhance SOC, through a $\Delta g$-mechanism (*21–23*), or due to hyperfine interaction (HFI) with paramagnetic nuclei (protons, nitrogen, etc.). In a spin-resonance experiment, microwaves with a properly chosen energy (frequency) flip the spin orientation and connect the $m_s = \pm 1$ and $m_s = 0$ substates of a triplet exciplex, which, in turn, facilitates population transfer to an emissive singlet exciplex final state. However, the underlying spin physics is quite complex as the spin-spin interactions in such electron-hole pairs depend strongly on the molecular environment, as well as on the details of how these excited states are created, e.g., optically excited or by injection from electrodes. (*24, 25*). Furthermore, the strength of the spin-spin interaction determines the energy gap between the states $m_s = 0$ and $m_s = \pm 1$, the so-called zero-field splitting (ZFS), and in practice is given by the spatial distance. As expected, the radii of the e-h pairs in the disordered organic systems have a very broad distribution. Thus, the sought-after spectroscopic precision of EPR methods for manipulating spin states is compromised by the fact that the excited spin-pairs [charge transfer (CT) states, exciplexes] are broadly distributed in a heterogeneous mixture of molecules, leading to inhomogeneously broadened envelopes instead of discrete resonance transitions (spin flips) and complicating access to the parameters of the spin Hamiltonian.

Here, we analyze the role of spin-spin interactions in the electroluminescence (EL) emission of an OLED using the method of EL detected magnetic resonance (ELDMR). We were able to disentangle the complexity associated with a broad distribution of spin pairs and propose an original technique to select sub-ensembles of e-h pairs and reconstruct the details of the inhomogeneously broadened ELDMR envelope in this way. For this purpose, a two-frequency hole-burning experiment is used, which pumps and saturates the spin system at a fixed frequency during the second probe frequency sweep (*26–29*). This pump-probe method allows not only the separation and addressing of individual spin packets in the inhomogeneously broadened ELDMR spectrum, but also the observation of coherent population oscillations (CPOs), which are observed as spikes with a very narrow linewidth (*26–35*). The hole-burning effect is caused by decoupling of the triplet exciplex spin state from the heterogeneous molecular environment, while the CPO can be described as a two-level quantum system oscillating with the beat frequency between pump and probe. The observation of CPO implies highly spin-polarized exciplex triplet states, which are not at all expected in an electrically driven OLED, where injected charge carriers are generally not spin-correlated.

**Results**

**Spin-Spin Interactions in OLEDs**
The results presented in this work were obtained on OLEDs based on the donor:acceptor system m-MTDATA:BPhen (4,4′,4″-tris[phenyl(m-tolyl)amino]triphenylamine: 4,7 diphenyl-1,10-phenanthroline) in a device configuration of PEDOT:PSS (40 nm) / m-MTDATA (30 nm) / m-MTDATA : BPhen (70 nm, 1:1 ratio) / BPhen (30 nm) / Ca (5 nm) / Al (120 nm), as shown in **Figure 1a**. With the pure m-MTDATA and BPhen layers as hole and electron transport layers, PEDOT:PSS (poly(3,4-ethylenedioxythiophene) poly(styrenesulfonate)) was used as a hole injection layer and a mixed layer of m-MTDATA and BPhen was used as an emission layer. Typical *j-V*-EL and external quantum efficiency (EQE) curves are shown in Supporting Figure S1. The photoluminescence (PL) spectra of the pure materials are shown in Figure 1b. The EL spectrum of the device made from the mixed layer is red-shifted and much broader compared to the emission spectra of the pristine molecules. The former is due to the formation of a bound exciplex state at the interface of the two molecules, with the electron localized at the lowest unoccupied molecular orbital (LUMO) of the acceptor BPhen and the hole localized at the highest occupied molecular orbital (HOMO) of the donor m-MTDATA, as schematically shown in the inset of Figure 1b (*36–39*). The much larger width is probably due to the broad energetic distribution of the emitting states, which we address below.

The electron and the hole both have a spin of $S = ½$. If these particles form a bound exciton or an exciplex (which we consider equivalent to a CT state), the total spin will be either $S = 0$ or $S = 1$. The Hamilton operator describing the triplet state is according to (*40*):

$$\mathcal{H} = \underbrace{-S_a^T J S_b}_{exchange} + \underbrace{g\mu_B \vec{S}\vec{B}}_{e-Zeeman} + \underbrace{S^T D S}_{ZFS} + \underbrace{S^T A I}_{HFI} \tag{1}$$

The first interaction term describes the exchange interaction of the two spins $S_a, S_b$ through the orbital overlap of their wave functions, where the exchange integral *J* determines the energy gap $\Delta E_{ST}$ between singlet and triplet energy levels. The second term is the electron-Zeeman interaction with the static external magnetic field *B*, the Landé factor *g*, the Bohr magneton $\mu_B$, and the spin operator **S**. The third and fourth terms describe spin-spin interactions: the electron-spin ZFS and the electron-nuclear HFI, with their respective tensors **D** and **A**. Quadrupole interaction and nuclear Zeeman splitting are neglected in this work because their contribution is negligible. A schematic representation of ZFS and broadening due to HFI is shown in Figure 1d. The $m_s = 0$ and $m_s = \pm 1$ triplet states are split by the ZFS parameter *D* (in frequency units) (*41*):

$$D/h = -\frac{\mu_0 \mu_B^2}{4\pi h} \frac{g_a g_b}{r_{ab}^3} (3\cos^2\theta - 1) \tag{2}$$

Here, $\mu_0$ is the vacuum permeability, $g_a$ and $g_b$ are the Landé factors for spins a and b, and $\theta$ is the angle between the direction of the magnetic field and the radius vector $r_{ab}$ connecting the two spins. The HFI term is given by the interaction of the exciplex triplet with surrounding paramagnetic nuclei represented by the nuclear spin operator *I*. This leads to a further splitting of the $m_s = \pm 1$ sub-levels, where each nucleus contributes with $A_{0,i}$ and the HFI term in Equation 1 is extended by the summation over all nuclei. The total number $N_{tot}$ of HFI-sub-levels for *i* non-equivalent nuclei is then given by $N_{tot} = (2I+1)^i$. For *i* equivalent nuclei, several HFI-levels degenerate and the total number is given by $N_{eq} = (2i \cdot I+1)$ (*42*).

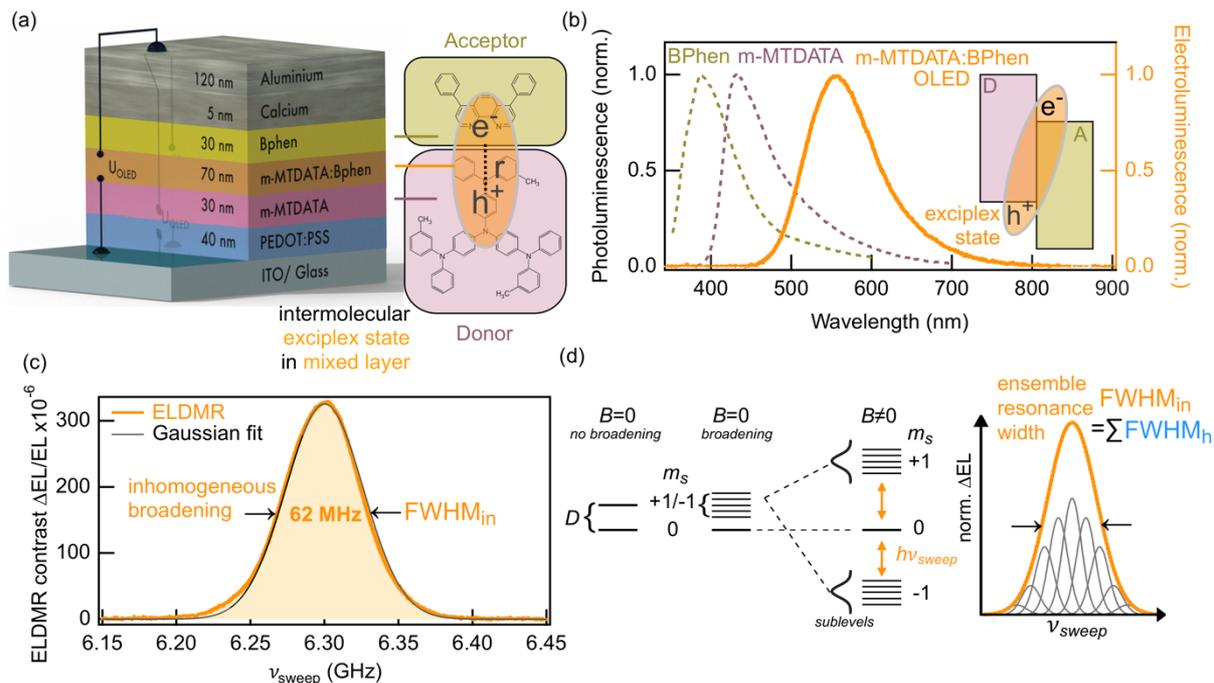

**Figure 1**. Magnetic resonance on OLEDs. (a) Schematic of the m-MTDATA:BPhen OLED architecture. (b) Photoluminescence spectra of pure materials (left axis, dashed lines) and electroluminescence (EL) spectrum of an OLED device (right axis, solid orange line). The inset shows a bound exciplex state formed at the donor:acceptor interface. (c) EL detected magnetic resonance (ELDMR) spectrum (orange line) of the m-MTDATA:BPhen OLED in a magnetic field of $B$ = 225 mT and a Gaussian fit (black line) with FWHM of 62 MHz. (d) Triplet sub-levels with zero-field splitting $D$ may exhibit broadening due to HFI. The sub-levels split in an external magnetic field and under resonant microwave irradiation with frequency $\nu_{\mathrm{sweep}}$, ΔEL is detected, which, as we will show later, is the envelope of the sum of homogeneous sub-spectra.

In conventional EPR, the external magnetic field is swept at a fixed microwave frequency, resulting in microwave absorption and spin-flip in the sample if the resonance criteria are met. For the ELDMR method used in this work, we couple an OLED to a stripline circuit that serves as an antenna for the frequency-swept microwaves at a fixed magnetic field. The strength of the effect is evaluated by the ELDMR contrast ΔEL/EL. A detailed description of ELDMR can be found in previous publications (*24, 25*), where also the singlet-triplet energy gap $\Delta E_{ST}$ = 58 meV was estimated for the MTDATA:BPhen blends, which is, by far, the largest energy contribution among the spin-spin interaction terms in Equation 1. An example ELDMR spectrum of an m-MTDATA:BPhen OLED device at $T$ = 220 K is shown in Figure 1c, with the microwave frequency $\nu_{\mathrm{sweep}}$ swept from 6.15 to 6.45 GHz at fixed magnetic field $B$ = 225 mT. The spectrum exhibits a featureless shape, which can be fitted with a Gaussian with a full width at half maximum (FWHM) of 62 MHz. This indicates that the ELDMR spectrum is inhomogeneously broadened. The working hypothesis that we verify in the following is that we are dealing with an ensemble of triplet states with different spin-spin interaction parameters and thus transition frequencies, showing up themselves as a featureless envelope, as schematically shown in Figure 1d. With knowledge of the details of the broadening of the ELDMR spectrum, e.g., due to the distribution of the spin-spin interaction, the broadening of the EL spectrum can also be better understood and potentially eliminated once it is assigned to a controllable structural or morphological parameter.

**Hole-Burning Spectroscopy**
To explore the origin of spectral broadening, we apply a two-frequency ELDMR technique realized by introducing a second frequency ($\nu_{\mathrm{pump}}$) fixed within the ELDMR spectrum while simultaneously sweeping the frequency $\nu_{\mathrm{sweep}}$ during the measurement, as schematically shown in **Figure 2a**. As illustrated, a dip occurs when a sufficiently high-power pump microwave field saturates a transition between triplet spin sub-levels (see Figures S2, S3). The width of the dip is associated to the energetic span of the sub-levels in the energy diagram (Figure 2a) (*26, 27, 30–35*). In Figure 2b, a continuous wave ELDMR spectrum is shown (orange curve). Applying the second microwave frequency results in a dip at the position of the applied pump frequency of $\nu_{\mathrm{pump}}$ = 6.3 GHz (Figure 2b, blue curve). This so-called "hole-burning" technique is widely used in optical absorption and fluorescence spectroscopy, as well as in spin resonance, but it has never been reported for ELDMR (*28, 29, 43–45*).

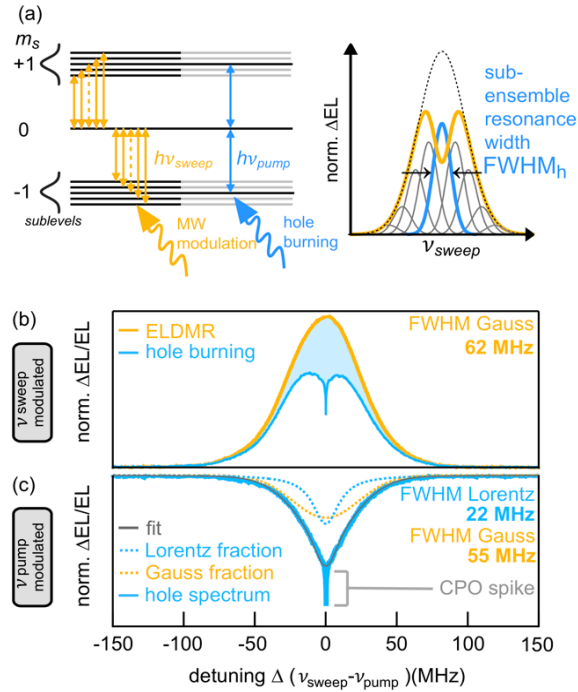

**Figure 2**. ELDMR hole-burning. (a) Schematic of the energetically broadened triplet sub-levels with resonance transitions induced by a variable (orange arrows) and fixed microwave (MW) frequency (blue arrows) and hole-burning in an inhomogeneously broadened ensemble. (b) ELDMR spectrum (orange) and hole-burning spectrum (blue) of m-MTDATA:BPhen for a frequency sweep of ±150 MHz around $\nu_{pump}$ = 6.3 GHz. The sweep frequency $\nu_{sweep}$ is modulated "on-off" for lock-in detection, and the pump frequency $\nu_{pump}$ is kept in continuous wave mode. (c) Directly measured hole spectrum by modulating the fixed frequency ($\nu_{pump}$) instead of the sweep frequency ($\nu_{probe}$). The signal consists of a broad Gaussian background (FWHM of 55 MHz), a narrower Lorentzian peak (FWHM of 22 MHz), and a very narrow spike in the center (FWHM of 12 kHz). The fitting curve (black) is a superposition of a Gaussian (dotted orange) and a Lorentzian (dotted blue) contribution.

To directly unveil the shape of the "spectral hole" shown in Figure 2b, we now modulate the pump frequency. The resulting spectrum in Figure 2c has a distinct, broad shape with a narrow spike on top. The latter is assigned to CPO, whose origin will be discussed in detail later. The broad spectrum can be perfectly fitted by the sum of Gauss and Lorentz functions with FWHM linewidths of 55 and 22 MHz, respectively, represented by the dotted lines in Figure 2c. Note that the "hole" spectral shape is independent of which of the two frequencies is on/off modulated, $\nu_{sweep}$ or $\nu_{pump}$, but because the pump frequency directly saturates a particular transition in the triplet sub-ensemble, we have used it in Figure 2c and in the following.

To investigate the structure of the hole in more detail, we varied the pump frequency, as shown in **Figure 3**. It shows a two-dimensional map in which the hole spectra are recorded as a function of $\nu_{pump}$ and $\nu_{sweep}$ in a frequency range from 6.25 to 6.35 GHz. The analysis shows that the hole spectrum can be decomposed into two contributions, Gaussian and Lorentzian. Furthermore, the Lorentzian contribution can be moved through the entire ELDMR spectrum, while the Gaussian contribution remains constant. The individual fits to the hole spectra are shown in Figure S4. Figure 3b exemplarily shows the corresponding cuts for three pump frequencies. When the pump frequency is shifted from the central $\nu_{pump}$ = 6.3 GHz to the side flanks, the holes become asymmetric, also indicating that the Gaussian background does not depend on $\nu_{pump}$ (Figure 3a, bottom left), but the Lorentzian line does (Figure 3a, bottom right). However, the width of the Gaussian background (55 MHz) is slightly less than the total measured ELDMR linewidth, which is 62 MHz, which could indicate a different origin. To verify this, we evaluated the two contributions separately and used them to reconstruct the experimental ELDMR spectrum. The individual linewidths for Gaussian and Lorentzian contributions are shown in Figure 3c. While the Lorentz linewidth remains constant at 22 MHz and does not change with $\nu_{pump}$, the Gauss linewidth varies between 50 and 62 MHz in the applied range of pump frequencies. As shown in Figure 3d, we can perfectly reconstruct the ELDMR spectrum by summing all contributions in the pump frequency range studied. This illustrates that the inhomogeneously broadened spectrum is the envelope of superimposed sub-spectra, that can be revealed by hole-burning spectroscopy.

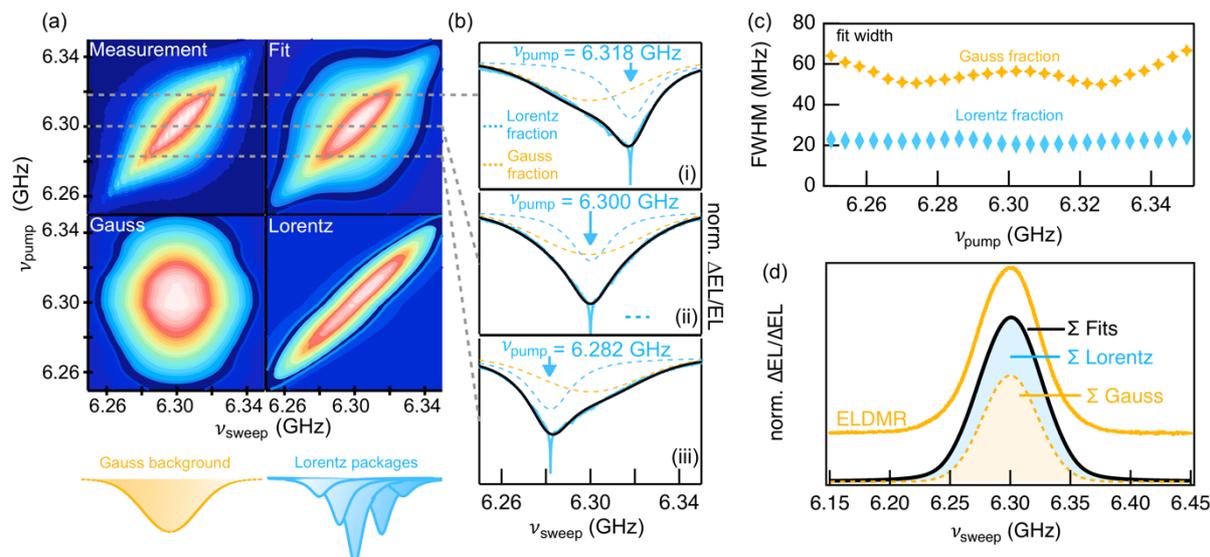

**Figure 3.** Spectral analysis of the hole by varying the pump frequency $\nu_{\text{pump}}$. (a) Color map of the hole spectrum as a function of $\nu_{\text{sweep}}$ and $\nu_{\text{pump}}$. The fit map is the sum of the Gauss and Lorentz maps. The Gaussian contribution is centered at $\nu_{\text{sweep}}$ = 6.30 GHz, while the Lorentzian varies with $\nu_{\text{pump}}$, as shown schematically at the bottom. (b) Hole contributions as a function of $\nu_{\text{pump}}$. While the Lorenz line shifts with pump frequency, the Gaussian contribution remains fixed. (c) FWHM for Gauss and Lorentz contributions as a function of $\nu_{\text{pump}}$. (d) Reconstruction of the ELDMR signal from the Lorentzian and Gaussian contributions. The sum of all fits has the same linewidth and shape as the ELDMR spectrum.

The fitted intensities of the Lorentz fraction can be further used to determine the dipolar interaction $D$ of electron and hole within the exciplex as shown in Figure S5b of the Supporting Information. The fitted intensities reveal the underlying triplet spectrum without the Gaussian background and this spectrum can be simulated with $D$ = 50±5 MHz. From $D$ we can also estimate the e-h separation within the exciplex (*41*) to be $r_{e-h}[\text{nm}] = \sqrt[3]{78.05/D[\text{MHz}]}$ nm = 1.1 to 1.2 nm, i.e., matching the expected delocalization over neighboring molecules. The coupling distance of emissive exciplex states is therefore quite uniform.

This analysis shows that the inhomogeneous broadening of the ELDMR spectrum is not due to a broad distribution of triplet exciplex states with different e-h separations and thus dipolar interactions. Instead, we propose spectral diffusion of overlapping spin packets as a reason for unresolved HFI, which has also been reported for other spin systems, e.g., for color centers in diamonds (*27*), and which we discuss in the following. The same process is responsible for the inhomogeneous broadening and thus for the Gaussian background.

**Inhomogeneous Spectral Broadening by Unresolved HFI**
With the hole-burning method, we revealed an underlying homogeneous linewidth and identified an additional inhomogeneous Gaussian broadening. For this broadening, we can exclude effects due to high microwave power and the $\Delta g$-mechanism as shown in detail in the Supporting Information. In the following, we will show that it can be assigned to unresolved HFI with paramagnetic nuclei in the molecules. For m-MTDATA:BPhen, this includes $i_H$ = 64 $^1$H protons with a nuclear spin of $I$ = ½ and $i_N$ = 6 $^{14}$N nitrogen nuclei with $I$ = 1, while other paramagnetic isotopes have negligible abundance. The isotropic hyperfine coupling constant $A$ depends on the electron-nuclei wavefunction overlap, the Fermi contact interaction (*46*). In organic molecular systems, e-h wavefunctions delocalize, involving more interacting nuclei, but decreasing the average HFI (motional narrowing) (*47–49*). Electron-proton HFI $A_H$ is usually weak and unresolved, leading to inhomogeneous broadening (*50–54*). For exciplexes in a similar TADF system, $A_{H,max}$ < 5.5 MHz was determined as an upper limit (*55*), whereas for charge-separated states in polymer:fullerene blends, $A_{H,max}$ of 2 to 3 MHz (*50*) or 4 to 5 MHz (*51*) was estimated. Considering the large molecular extent of m-MTDATA:BPhen and that protons at the edges of molecules are not part of the conjugated electron system, we can expect $A_H$ < 2 MHz. HFI with a discrete number ($i_N$ = 4) of $^{14}$N nuclei directly incorporated in the conjugated system of triphenylamine-derivatives has been determined to be $A_N$ = 2.5 to 3.8 MHz (*56*). We performed exemplary HFI simulations (*80*) of the exciplex triplet with $i_N$ = 6 $^{14}$N as presented in the Supporting Information and estimate $A_N$ < 2 MHz.
The unresolved and weak HFI in m-MTDATA:BPhen can explain the inhomogeneous linewidth broadening. More intriguing, however, is the fact that $A_{H,N}$ < 2 MHz is considerably smaller than the ZFS $D < 50$ MHz. This suggests that HFI is too weak to mediate RISC rates as the exciplex triplet is decoupled from the nuclear spin bath by the stronger ZFS $D$.

**Coherent Population Oscillations of Spin-Polarized Exciplex States**

In a hole-burning experiment, we apply two microwave fields to the spin system. From a classical viewpoint, when two electromagnetic waves with nearby frequencies interfere, a beat oscillation of intensity occurs with the beat frequency $\Delta\nu = \nu_{\text{sweep}} - \nu_{\text{pump}}$, where $\Delta\nu$ is the detuning between pump and probe frequencies (see Figure S9). While the application of a single microwave resonance frequency tends to equilibrate the population of spin sub-levels, the simultaneous application of two frequencies to a highly spin polarized system can cause the triplet population to oscillate, especially when the beat frequency is in the range of or lower than the inverse spin relaxation time of the spin system. This CPO effect has been reported for various inorganic spin systems in the past (*27, 44, 57, 58*) and is also known from laser spectroscopy of two-level systems (*59*). In a lock-in detected ELDMR experiment, as in ours, these oscillations are time-averaged. This leads to a beat frequency $\Delta\nu$-dependent depth of the burned spectral hole. Spin relaxation is mediated by a longitudinal spin-lattice relaxation time $T_1$ and a transverse dephasing time $T_2^*$, and they will determine the eigenfrequencies of the spin state. In the case of a forced oscillation with the variable beat frequency, the "hole" depth will be affected. In a continuous detuning scan, the beat frequency is varied from a few Hz to MHz, and we observe such a very narrow spike on the top of the hole, as shown in **Figure 4**. The homogeneous linewidth $\text{FWHM}_h = 22$ MHz allows to estimate a lower limit for the dephasing time $T_2^* > 2/(\pi \cdot \text{FWHM}_h)$ = 30 ns, while the width of the spike corresponds to the spin-lattice relaxation time. The dephasing time agrees well with that estimated from pulsed Rabi experiments on similar OLED systems (*55, 60*).

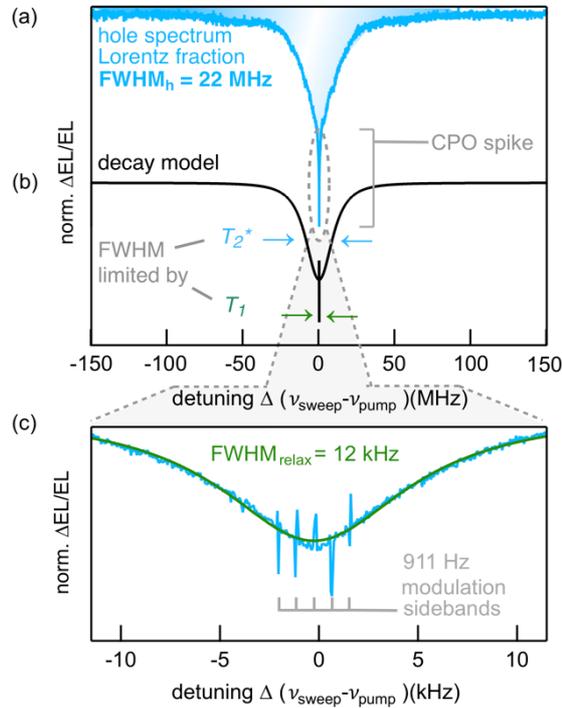

**Figure 4**. Coherent Population Oscillations. (a) Lorentz fraction of the hole spectrum with CPO spike. We subtracted the Gaussian background (see Fig. 3a) for clarity. (b) FFT of the decay model in Equation 3. The FWHM is limited by dephasing time $T_2^*$ and spin-lattice relaxation time $T_1$ (c) Zoomed-in measurement of the hole spectrum. The spike can be described by one Lorentzian, which results from CPO between triplet levels due to microwave beating between $\nu_{\text{sweep}}$ and $\nu_{\text{pump}}$.

When changing from frequency to time domain, one can assume a simple decay model. The net magnetization of a sub-ensemble of $S = 1$ states formed at $t = 0$ decays exponentially with the time constants $T_2^*$ and $T_1$. In an external magnetic field, it is also undergoing precession with the Larmor frequency $\nu_L = \omega_L/2\pi$. Thus, we can write for the intensity of net magnetization of a sub-ensemble:

$$I(t) = \left[A \cdot \exp\left(-\frac{t}{T_2^*}\right) + B \cdot \exp\left(-\frac{t}{T_1}\right)\right] \cdot \sin(\omega_L \cdot t), \tag{3}$$

with amplitudes *A* and *B*, respectively. The fast Fourier transform (FFT) of this equation is shown in Figure 4b. The fast decay due to $T_2^*$ gives a broad signal, while the FFT of the long decay with $T_1$ reproduces the spike. By zooming in on the spike in Figure 4a, it turns out that the shape is more complex and cannot be perfectly fitted with a single Lorentzian function. The shape can however be fully reproduced by applying a modified three-level model of (*27*). This is possibly due to inhomogeneous broadening of the triplet sub-levels involved, but the issue cannot be fully elucidated. For further details, we refer to the Supporting Information. We can nevertheless estimate FWHM$_{relax}$ = 12 kHz for the narrowest contribution as shown in Figure 4c with which we can estimate an upper limit of the spin-lattice relaxation time $T_1 < 2/(\pi \cdot \text{FWHM}_{relax})$ = 50 μs. These measurements were performed at an experimental temperature of *T* = 220 K. In between 290 and 160 K (Figure S12) we observed that *T$_1$* is not very temperature-sensitive as it increases from 50 to 70 μs when cooling. Modulation sidebands appear due to on-off modulation of the pump frequency with $\nu_{mod}$ = 911 Hz, which rules out the possibility that on-off modulation effects are responsible for contributions to CPO in the kHz to MHz range. We also exclude $\nu_{mod}$ to influence the determined *T$_1$* time by measuring with different modulation frequencies (see Figure S13).

**Discussion**

**Interplay Between RISC and Spin Relaxation of Triplet Exciplex States**
In the following, we discuss the role of the determined spin-spin interactions and spin relaxation times in comparison with RISC time constants. Interaction strengths (*A*, *D*, and $g\mu_B B$) and FWHM linewidths are given in frequency units, while time constants are always given in time units and not as rates.

The observation of CPO in the EL of working OLEDs is notable as it proves that the emitting exciplex states are highly spin-polarized. This is a remarkable finding as injected charge carriers are generally not spin-correlated and, upon recombination at the donor:acceptor interface, will populate the exciplex singlet and triplet $m_s = 0, \pm 1$ sub-levels evenly. Singlet states recombine on a nanosecond time scale, leaving the triplets to undergo RISC to facilitate delayed fluorescence. The fact that we observe spin-polarized exciplexes in electrically driven devices unambiguously teaches us that recombination of triplet exciplexes is highly spin-dependent with unequal RISC rates for the $m_s = 0, \pm 1$ sub-levels. The selection rules ($\Delta S = \pm 1$, $\Delta m_s = 0$) for transitions between singlet (*S* = 0, $m_s = 0$) and triplet (*S* = 1, $m_s = 0, \pm 1$) exciplex states would favor faster depletion of $m_s = 0$ triplet exciplexes and can thus be responsible (compare Figure S6).

Spin polarization decays with the spin-lattice relaxation time *T$_1$* < 50 μs, which by far exceeds RISC on the timescale of 30 to 220 ns for this material system (*25*). Similar *T$_1$* times around 30 μs were reported for other OLED systems and even organic solar cells probed by electrically and optically detected magnetic resonance (EDMR, ODMR) (*61–64*). A long spin relaxation time is hereby typical for organic materials with weak HFI and strong ZFS as this decouples the electron spin system from relaxation caused by the nuclear spin bath. Note that while the presented experiments are carried out in an external magnetic field that results in a triplet state Zeeman splitting of $g\mu_B B$ = 6.3 GHz, i.e., much larger than ZFS and HFI, the spin polarization mechanism will be identical even without applied magnetic field. The ZFS of *D* = 50 MHz is sufficiently large to decouple the electron spin system from the nuclear spin bath in zero magnetic field, preserving a long *T$_1$* time. RISC time constants for other donor:acceptor exciplex (*9*, *65–68*) and molecular TADF materials (*69–72*) were reported around 0.5 to 2 μs (*67–71*) or up to 30 μs (*65*, *66*, *72*), always being shorter or even significantly shorter than *T$_1$*. This imbalance between RISC and *T$_1$* time constants by up to more than an order of magnitude leads to spin polarization between $m_s = 0, \pm 1$ levels, and we can thus assume it to be in the order of or greater than 90% ($= 1 - \frac{T_{RISC}}{T_1}$). Therefore, in most TADF OLEDs, a strong spin polarization is expected, which unnecessarily prolongs the lifetime of the excited states. This may ultimately limit quantum efficiencies, especially at high OLED current densities, and lead to efficiency roll-off that is usually ascribed to triplet-triplet and triplet-charge annihilation (*10*, *12*, *66*). This finding strongly implies that spin polarization of triplets must be suppressed by a mechanism to accelerate spin relaxation. *T$_1$* can be substantially shortened by molecular curvature-enhanced SOC, as has been shown theoretically (*73*) and experimentally (*74*, *75*). In addition, *T$_1$* scales with the $\Delta g$-mechanism as $T_1 \sim (\Delta g)^{-2}$ (*76*). Both mechanisms can be used for the design of next-generation TADF systems.

In contrast to the long *T$_1$* time, the spin dephasing time *T$_2$*\* > 30 ns is exceptionally short and actually limits the homogeneous linewidth FWHM$_h$ = 22 MHz. Such fast dephasing is often mediated by interaction with multiple, almost equally coupled nuclei (*77*), as in our case with nitrogen nuclei and protons. RISC is on the same timescale or even considerably longer than *T$_2$*\*, which means that spins will dephase quickly and RISC is not affected by a short *T$_2$*\*. Consequently, the ensemble of exciplexes in an OLED is highly incoherent, i.e., *T$_2$*\* is short enough to not affect the operation of the OLED at all.

**Comparison with other Donor:Acceptor Systems**

Last, it is intriguing to compare the exciplex states in intermolecular TADF emitters studied here with excited states in other material systems, such as CT states in intramolecular TADF emitters and polymer:fullerene blends used in organic photovoltaics (OPV) (while exciplex and CT states are essentially the same kind of species). Both exciplex- and CT-based TADF emitter systems exhibit high IQEs and have singlet-triplet gaps $\Delta E_{ST}$ in the range of thermal energy. However, dipolar interactions differ by at least an order of magnitude, with $D$ > 500 MHz determined for intramolecular TADF emitters (*18*, *20*), indicating their stronger localization compared to the system studied here. In contrast, OPV CT states have an order of magnitude smaller dipolar interaction of $D$ < 5 MHz and $\Delta E_{ST}$ < 5 MHz, which corresponds to 20 neV, i.e., more than six orders of magnitude smaller than for TADF (*78*). This corresponds to a much wider excited state delocalization and a weaker oscillator strength. Hence, it is not unexpected that OPV CT states exhibit very low EL quantum efficiency in the order of $10^{-6}$ (*79*). We conclude that the m-MTDATA:BPhen exciplex emitter with $D$ = 50±5 MHz and a singlet-triplet gap of $\Delta E_{ST}$ = 58 meV (*24*, *25*) is an intermediate case, between highly emissive intramolecular TADF and poorly-emissive OPV CT states, revealing large or nearly zero dipolar interactions, respectively.

**Conclusion**

In conclusion, we have experimentally shown that electrical injection of charge carriers into OLEDs based on the TADF system m-MTDATA:BPhen results in the buildup of a spin-polarized triplet exciplex population with long spin-lattice relaxation time. This phenomenon is initially due to spin-statistically injected charges in the devices but can be considered detrimental to OLED efficiency because these exciplex species are not undergoing RISC. The observation was made possible by implementing a novel two-frequency spin-resonance spectroscopy based on the so-called "hole-burning" by direct monitoring of the EL intensity. We used it to observe coherent population oscillations that exhibit an electroluminescence spike of extremely narrow width, from which we not only prove >90% spin-polarization itself but also estimate the spin-lattice relaxation time of 50 μs, which is thus longer than the RISC time constants in this and similar TADF systems. Furthermore, our spin-resonance protocol provided insights into the inhomogeneously broadened triplet exciplex spectra consisting of individual sub-ensembles or "spectral holes" with a characteristic ZFS parameter $D$ = 50±5 MHz, indicating exciplex states of 1.1 to 1.2 nm extent, i.e., over adjacent donor:acceptor molecular pairs. The broadening is the result of the disordered nature of the molecular blends, but it can also be quantitatively understood as the result of unresolved HFI with the surrounding protons and the nitrogen nuclei, and we provide an estimate of the strength of an averaged nuclear field that is in the range of 2 MHz. The fact that the spin relaxation time is much longer than the RISC time clearly implies that the type of TADF OLEDs studied here is efficiency-limited due to the spin polarization, which needs to be suppressed by a mechanism to accelerate the relaxation, such as SOC. Our findings further put the commonly targeted small singlet-triplet gap and thus high RISC rate into perspective.

**Materials and Methods**

**Device Fabrication**

For OLED devices, ITO-covered glass substrates (Vision-Tek Systems) were used. The hole injection layer PEDOT:PSS (4083Ai, Heraeus) was spin coated at 3000 rpm, resulting in a 40 nm thick film. It is thermally annealed for 10 min at 130°C. Remaining layers are evaporated under vacuum. The hole transport layer (m-MTDATA) and electron transport layer (BPhen) have a thickness of 30 nm. In-between, a 70 nm mixed layer of both materials is deposited as emission layer. A top electrode consisting of calcium (5 nm) and aluminum (120 nm) was used, resulting in an active area of 3 mm$^2$.

**Device Characterization**

*J*-*V*-EL and EQE characteristics were measured with a parameter analyzer (Agilent 4155C), and a silicon photodetector was placed on top of the OLED.

**Magnetic Resonance**

For ELDMR measurements, we used a modified EPR spectrometer (Bruker ESP300) with a nitrogen flow cryostat (Oxford 935). The temperature for all measurements was $T$ = 220 K, except for Figure S12. The microwaves are generated with a signal generator (Anritsu MG3694C). The output power for all measurements was set to 0 dBm, amplified by a 3 W amplifier (Mini-Circuits ZVW-3W-183+). The OLED device was placed on top of a microwave strip-line to enable spin control in the emission layer. A source measure unit (Keithley 237) drives the OLED in constant current mode ($j$ = 1.6 mA/cm$^2$), and the EL is detected with a photodiode (Hamamatsu, S2387-66R) and amplified with a current-voltage amplifier (Femto DLPCA-200). The change of EL upon resonant microwaves, ΔEL, was detected with a lock-in-amplifier (SR7265).

**Hole-Burning Spectroscopy**
In a hole-burning experiment, the pump microwave generator (Synth HD, Windfreak) was amplified with a secondary 3W amplifier. The outputs of both amplifiers are merged with a high-power combiner (Mini-Circuits ZN2PD-183W-S+). Hole-burning spectra are detected by modulation of the sweep microwave generator. Direct detection of hole spectra is realized by modulation of the pump generator.

**Supplementary Materials**
Device characteristics, spectral broadening mechanisms including CPO simulations, dependence of $T_1$ time on experimental conditions.

**Acknowledgements**
We thank Jeannine Grüne for fruitful discussions. This publication was supported by the Open Access Publication Fund of the University of Würzburg. S.W. acknowledges support by the German Research Foundation, DFG, within FOR1809 (DY18/12-2). A.G., V.D. and A.S. acknowledge financial support from the DFG through the Würzburg-Dresden Cluster of Excellence on Complexity and Topology in Quantum Matter – ct.qmat (EXC 2147, project-id 39085490). **Author Contributions:** S.W., A.G. and A.S. designed the experiments that were performed by S.W. and R.B. Data analysis was performed by S.W., A.G. and A.S. The draft was written by S.W., V.D. and A.S., with all authors contributing in iterations. The project was supervised by V.D. and A.S. **Competing Interests:** The authors declare that they have no competing interests. **Data Availability:** All data needed to evaluate the conclusions in the paper are present in the paper and/or the Supplementary Materials.

**ORCID**
Sebastian Weissenseel: 0000-0001-9811-1005
Andreas Gottscholl: 0000-0002-9724-1657
Vladimir Dyakonov: 0000-0001-8725-9573
Andreas Sperlich: 0000-0002-0850-6757

# Supporting Information

**Long-Lived Spin-Polarized Intermolecular Exciplex States in Thermally Activated Delayed Fluorescence-Based Organic Light-Emitting Diodes**

Sebastian Weissenseel, Andreas Gottscholl, Rebecca Bönnighausen,
Vladimir Dyakonov, Andreas Sperlich*
* sperlich@physik.uni-wuerzburg.de

Experimental Physics 6 and Würzburg-Dresden Cluster of Excellence ct.qmat, Julius Maximilian University of Würzburg, 97074 Würzburg, Germany

**Supporting Information Contents**
- Device Characteristics
- Microwave Power Broadening
- Lorentzian and Gaussian Contributions in Hole Burning Spectra
- Spectral Broadening due to Dipolar Interactions
- Spectral Broadening due to Hyperfine Interactions
- Simulation of Hyperfine Interactions
- Spectral Broadening due to $\Delta g$-Mechanism
- Microwave Beating by Application of Two Microwave Fields
- Simulation of Coherent Population Oscillation (CPO)
- Temperature-Dependence of $T_1$

**Device Characteristics**

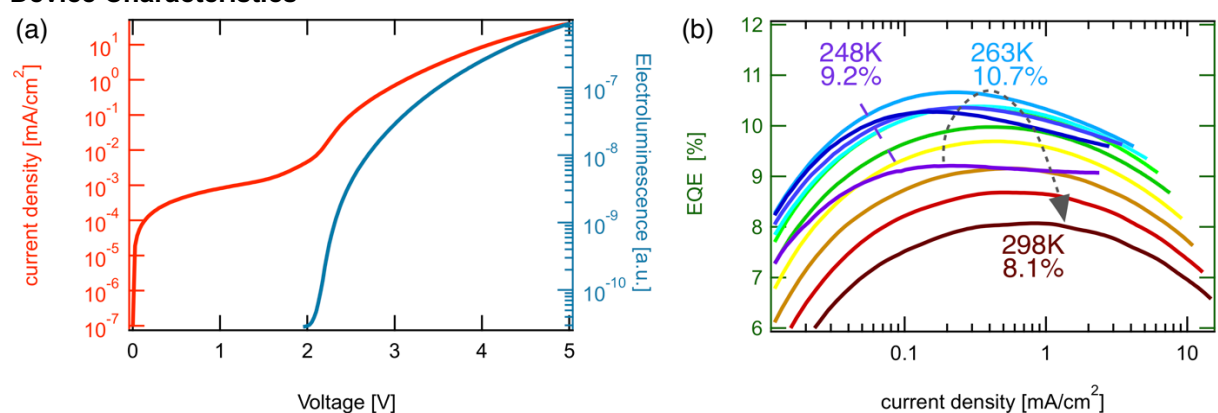

**Figure S1.** OLED performance characteristics. (a) *j-V*-EL and (b) EQE characteristics of a typical m-MTDATA:BPhen OLED. Measurements in (b) were performed at 248 – 298 K with a maximum EQE of 10.7% at *T* = 263 K.

The *j-V*-EL curve shows a linear increase of the current density due to ohmic parallel resistance at low voltage. Above 2 V, the exponential increase of the current density causes exponential increase of EL emission (onset of EL at 2 V). Above 3 V, ohmic series resistance is dominating, which decreases the slope in current and EL. ELDMR measurements have been performed in constant current mode (*j* = 1.6 mA/cm$^2$). The EQE-*j* curves show a maximum EQE of 10.7% at 263 K and somewhat smaller values for higher and lower temperatures. The efficiency roll-off is visible but not severe in the tested current density range.

## Microwave Power Broadening

If a spin system is pumped by high microwave powers, the intensity of magnetic resonance signals is driven to saturation. In this case, the linewidth can increase due to microwave power broadening. A power series of the hole burning experiments is shown in Figures S2 and S3. We tested the influence of the hole burning pump power in two different measurement modes: lock-in detection with on/off-modulation of $\nu_{\text{sweep}}$ and, secondly, on/off-modulation of $\nu_{\text{pump}}$. In the first case in Figure S2, by decreasing the microwave power of a maximum $P_{\text{pump}} = 3\,\text{W}$ to 1%, the hole burning spectrum approaches the ELDMR spectrum. A detailed view is given in the zoomed-in measurement in Figure S2b. In contrast, by detection via pump modulation in Figure S3, both, the Lorentzian and the Gaussian contributions can be evaluated. The intensity of both contributions is still in the linear regime. Thus, with the hole burning experiment we do not saturate the triplet sub-resonances so extensively that the spectra would broaden (Figure S3b). Further, by variation of the pump power over two orders of magnitude, no change in linewidth is observed (Figure S3c). Hence, we observe that the full-width ELDMR spectrum and the hole burning resonances are not broadened due to high microwave power.

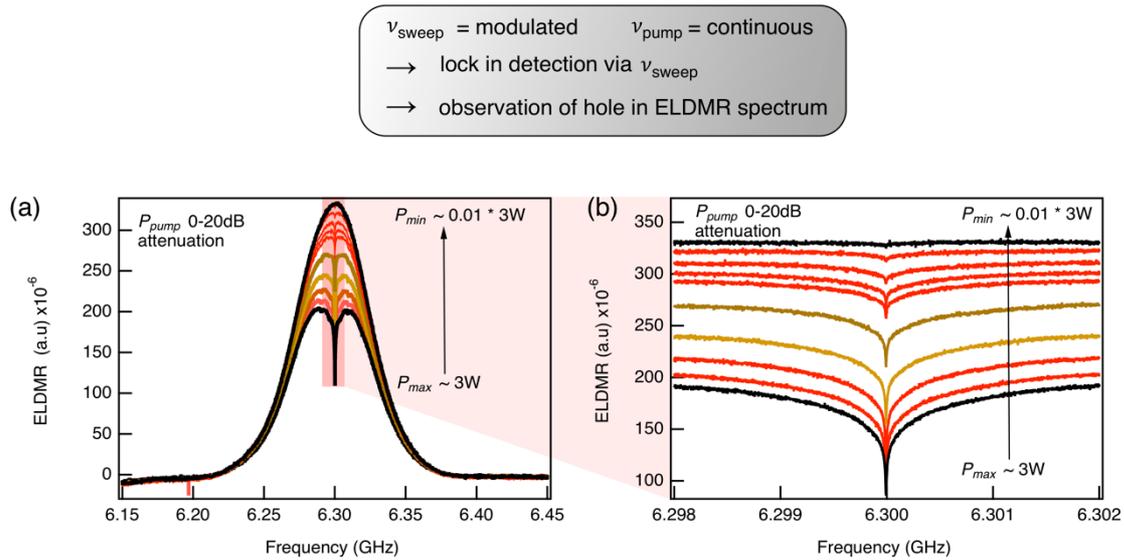

**Figure S2:** Microwave power dependence of the hole burning spectrum with on/off modulation of $\nu_{\text{sweep}}$. (a) With vanishing microwave power of $P_{\text{pump}}$, the hole burning spectrum approaches the ELDMR spectrum. (b) Zoomed-in measurement of the hole.

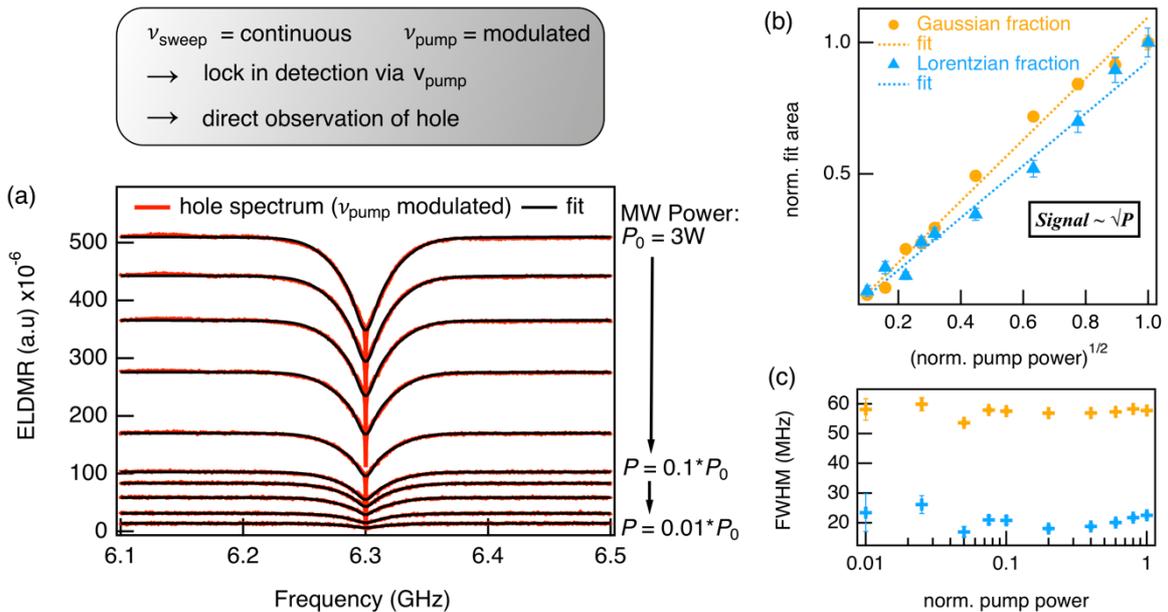

**Figure S3:** Microwave power dependence of the hole burning spectrum with on/off modulation of $\nu_{\text{pump}}$. (a) The hole burning spectrum decreases with $P_{\text{pump}}$. (b) The Hole Intensity is proportional to the square root of the pump microwave power. There is no visible saturation of the intensity. (c) There is no change in the linewidth of the spectrum, which excludes power broadening.

**Lorentzian and Gaussian Contributions in Hole Burning Spectra**

Here, we show the respective fits, which are shown in the two-dimensional map in Figure 3 in the main text. The Lorentzian contributions (Figure S4a) are shifting with the pump frequency. The corresponding gaussian contributions (Figure S4b) remain at the centered resonance frequency. The summed fit is shown in Figure S4c. If a hole is burned at the edge of the ELDMR spectrum this results in a highly asymmetric hole spectrum (Figure S4d). The dashed line shows the Gaussian (centered) and the Lorentzian contribution of the fit (black curve) of a hole burned at $\nu_{\text{pump}}$ = 6.282 GHz.

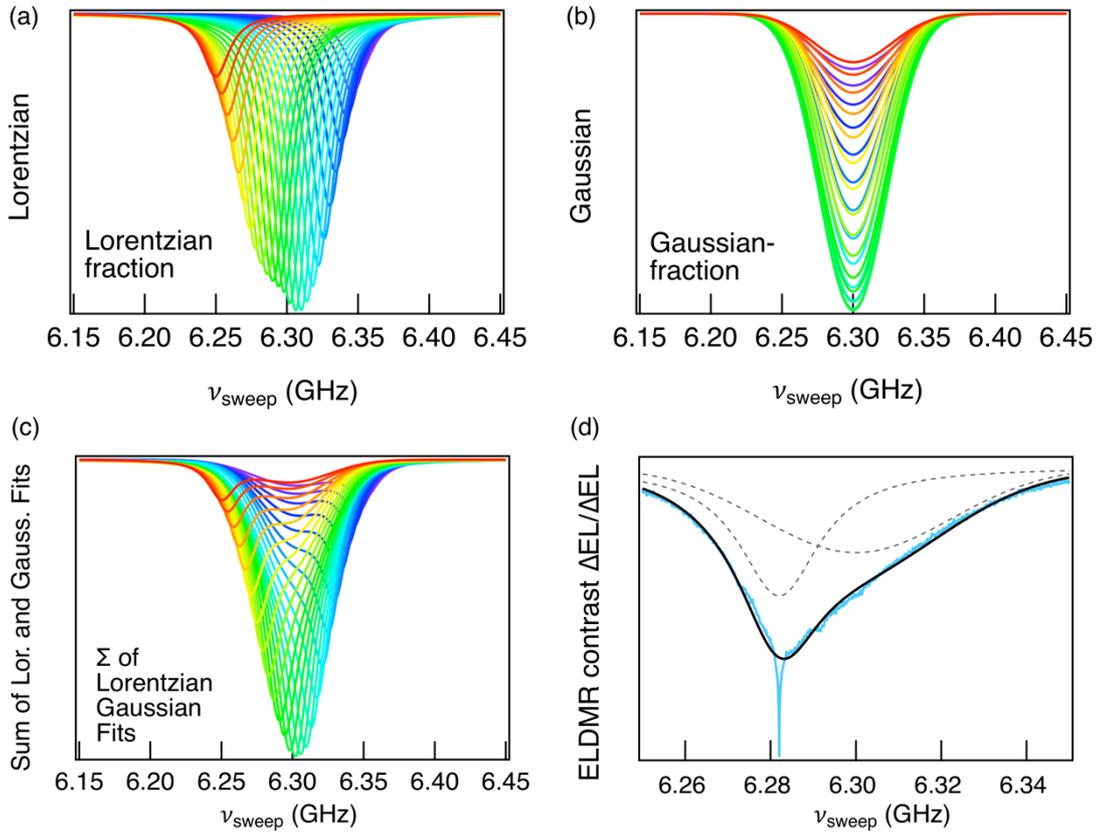

**Figure S4.** Hole burning at different pump frequency positions varying $\nu_{\text{pump}}$ from 6.25 GHz to 6.35 GHz. (a) and (b) show the Gaussian and Lorentzian fractions of the summed fit in (c). The center of the Gaussian fit remains at the center frequency of 6.3 GHz. The center of the Lorentzian fit is kept at the pump frequency. (d) Exemplary fit to a hole spectrum with pump frequency $\nu_{\text{pump}}$ = 6.282 GHz, dashed lines are the Lorentzian (dashed left) and the Gaussian (dashed centered) contributions.

**Spectral Broadening due to Dipolar Interactions**

Dipolar interactions of positive and negative charge within spin triplets ($S = 1$, $D > 0$) lead to broad, characteristic magnetic resonance spectra with two intense peaks and two flat shoulders – so-called powder patterns. The spectral separation of the peak positions is $D$ and the shoulders are separated by $2D$ as shown in the following simulations using the matlab tool EasySpin.
The simulations are based on the following spin-system:

```
Sys0.S = 1;      % specifies the spin system to be a S = 1 particle
Sys0.g = [2.002]; % sets the g-factor
Sys0.D = [D];    % sets the dipolar interaction D
Sys0.lw = [0 5]; % sets the Lorentzian linewidth to 5 MHz
```

The spectrum is then simulated with the EasySpin function pepper. Simulations for different $D$ values are shown in Figure S5a and clearly show resolved sub-structures for higher $D$ values. Note, that the homogeneous linewidth is set to a Lorentzian linewidth of 5 MHz.

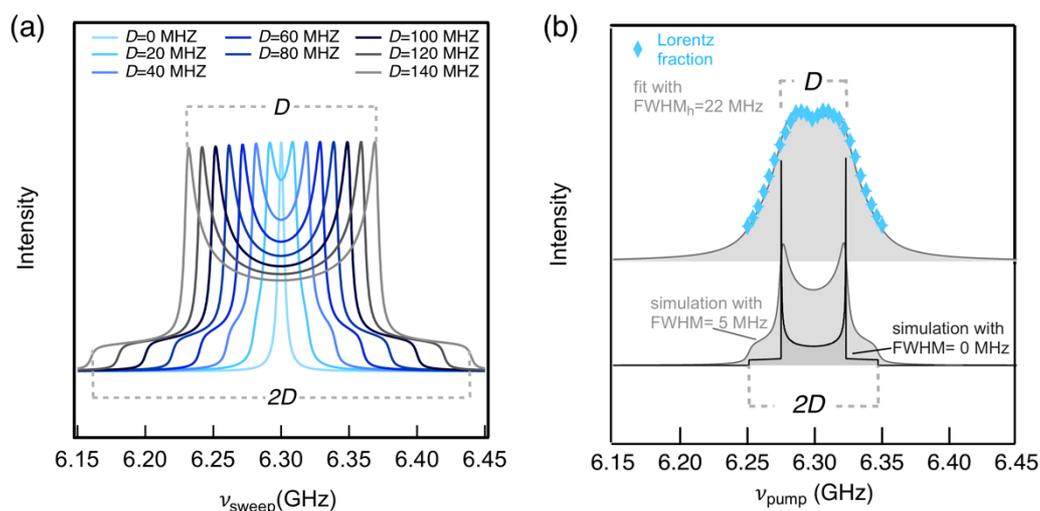

**Figure S5.** Simulation of triplet spectra. (a) The dipolar interaction parameter $D$ is varied in between 0 MHz and 140 MHz. The peaks are separated by $D$, while the lower-intensity shoulders are separated by $2D$. (b) The intensity of the Lorentzian fits of Figure S4 and Figure 3 of the main text (blue diamonds) exhibits two maxima. The data can be simulated (grey line) with a triplet powder pattern with $D = 50$ MHz (peak-to-peak separation) if a homogeneous linewidth of $FWHM_h = 22$ MHz is assumed. For comparison: the same triplet powder pattern is simulated with FWHM = 5 MHz and FWHM = 0 MHz (grey spectra).

The ELDMR spectra in Figures 2 and 3 of the main text are inhomogeneously broadened to an extent that no peaks and shoulders can be discerned to directly determine the dipolar interaction. Instead, we use the intensities of the Lorentzian fits in Figure S4a and plot them in Figure S5b. This method reveals the underlying triplet spectrum of two overlapping peaks without the Gaussian background. We use EasySpin to simulate a triplet powder pattern with $D = 50$ MHz without broadening (black trace), with FWHM = 5 MHz broadening (dark grey) and using the homogeneous hole linewidth of $FWHM_h = 22$ MHz. Remarkably, the experimental intensities can be reproduced extremely well with the latter and we derive $D = 50\pm5$ MHz for the dipolar interaction including simulation uncertainties. Thus, we revealed the hidden intensities of sub-ensembles in an inhomogeneously broadened spectrum.

**Spectral Broadening due to Hyperfine Interactions**

In the main text it is discussed that besides dipolar interactions, we expect the $^{14}$N HFI to add substantial ELDMR broadening. To derive an upper limit for $A_N$, we consider the ELDMR linewidth and the number of hyperfine levels resulting from interaction with $i_N = 6$ $^{14}$N nuclei, which are directly incorporated into the conjugated system. In the case of equal interaction of $^{14}$N nuclei, this leads to $n_{eq} = (2 \cdot i_N+1) = 13$ HFI-levels. We obtain an upper limit for $A_N$ by the following assumptions:

- Considering only $A_N$ and ignoring $A_H$, the width of the inhomogeneously broadened ELDMR spectrum is dominated by $D$ and $A_n$
- For $n_{eq} = 13$ HFI-levels, we expect $2i_N = 12$ splittings with separation $A_N$ (Figure S6).
- Comparing the difference of the lowest and highest frequency transitions:
    - the highest frequency is given by $g\mu_B B + D + \Sigma A_N I = g\mu_B B + D + 6A_N$
    - the lowest frequency is given by $g\mu_B B - D - \Sigma A_N I = g\mu_B B - D - 6A_N$
    - the difference of the two: $2D + 12A_N$
    - the shoulder-to-shoulder powder pattern width in Figure S5b is ~2× FWHM = 2× 62MHz

Consequently, we estimate the HFI strength for $i_N = 6$ $^{14}$N nuclei by the maximum shoulder separation and the difference between the highest and lowest transition frequency to

$$A_N \leq (2\ FWHM_{ELDMR} - 2\ D)/(2\ i_N) = 2\ MHz \tag{S1}$$

We can now derive the energy diagram in Figure S6. Without external magnetic field, the triplet states are split into $m_s = 0$ and $m_s = \pm 1$ sub-levels by a ZFS of 50 MHz. The Zeeman interaction of 6.3 GHz is opened by an external magnetic field of $B$ = 225 mT. HFI with one nitrogen nucleus ($^{14}$N, $I = 1$) further splits the $m_s = \pm 1$ triplet sub-levels into three levels each. The interaction with just one $^{14}$N-nucleus would results in $A_{N,i=1} \leq 12$ MHz, while additional nuclei generate further splittings and hence reduce $A_N$.

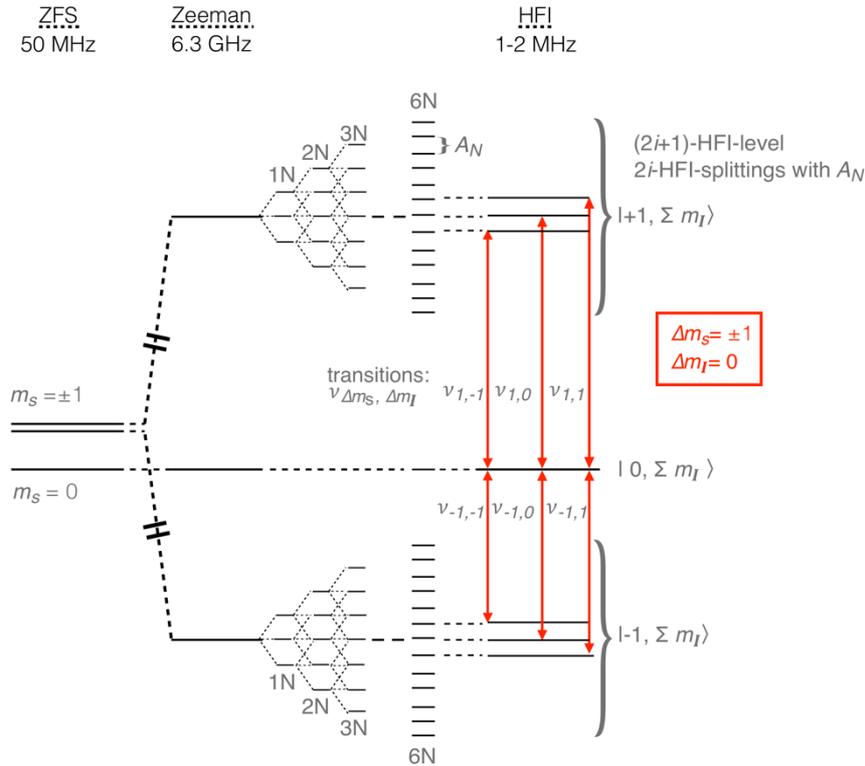

**Figure S6.** Energy diagram for a triplet exciplex including all relevant spin-spin interactions (not to scale). The ZFS $D$ = 50±5 MHz splits the $m_s = 0$ and $\pm 1$ states. At a resonant magnetic field of $B$ = 225 mT the $m_s = \pm 1$ levels split further into an upper and a lower triplet level by the Zeeman interaction of $g\mu_B B$ = 6.3 GHz. These levels possess an additional hyperfine structure depending on the number of interacting paramagnetic nuclei and their interaction strength $A$. Selection rules for transitions $\nu_{m_s,m_I}$ allow $\Delta m_s = \pm 1$ changes while maintain $\Delta m_I = 0$.

**Simulation of Hyperfine Interactions**

As a next step, we include hyperfine interactions into EasySpin simulations to explain the Gaussian broadening of the ELDMR spectrum and describe it by unresolved hyperfine interactions. The number of nuclei interacting with an m-MTDATA:BPhen exciplex will be determined by the $^{14}$N nuclei ($i_N$ = 6), directly incorporated into the conjugated system, and the surrounding protons $^{1}$H ($i_H$ = 64). The probability density of the exciplex wavefunction is more localized among the nitrogen nuclei than among the surrounding protons. Accordingly, the influence of the $^{14}$N on the linewidth predominates. In the following we thus discuss an example that considers only the influence of nitrogen. Slightly different parameters would still yield the same result: A distribution of hyperfine interactions lead to Gaussian line shapes with unresolved hyperfine interactions.

In Figure 2 (main text) we discussed that a hole burning experiment excites only a sub-ensemble of all available exciplex triplets. Exemplarily, here we assume a mean interaction of three nitrogen nuclei and omit the influence of the protons. Further, a distribution of $A_N$ due to disorder in the organic film is to be expected. The dihedral angles of the phenyl rings of the donor and acceptor molecules vary, which results in a change of geometry that can break the conjugation and prevent wavefunction delocalization over the full molecule. This again changes the spin density of the exciplex at the spatial coordinates of the nuclei and HFI strength $A_N$.

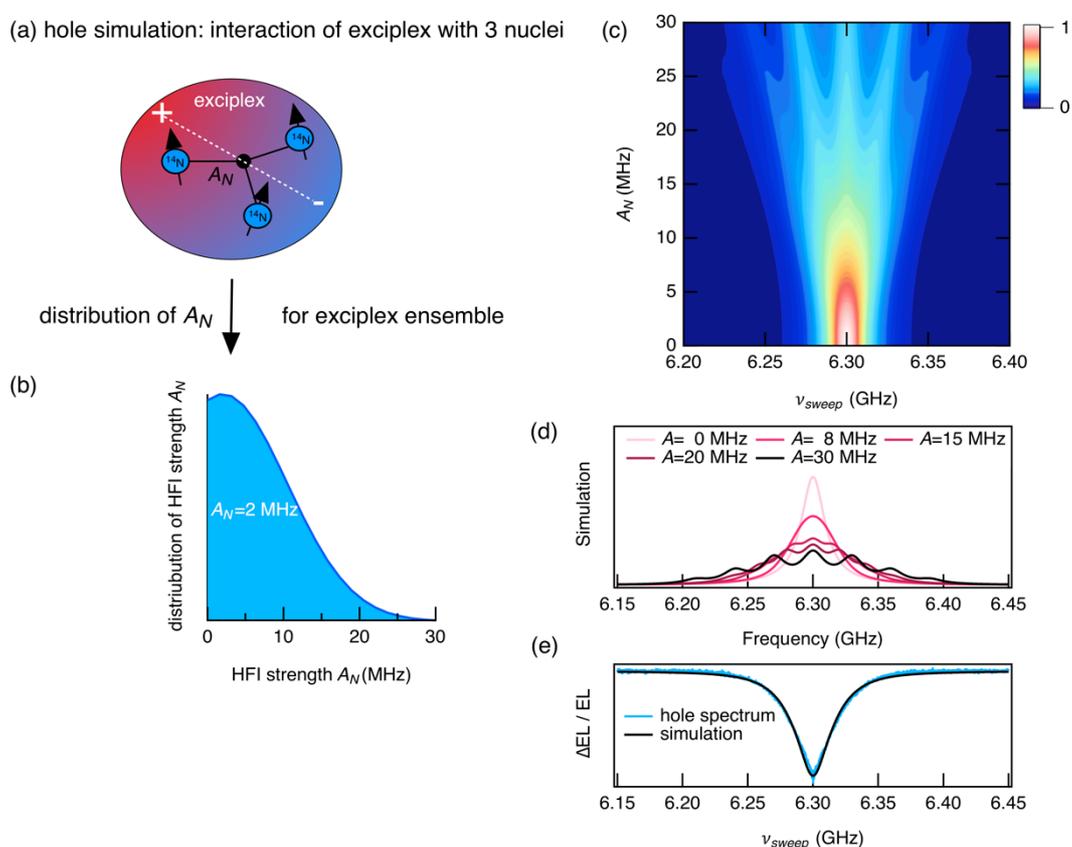

**Figure S7.** EasySpin simulation of a triplet sub-ensemble with interaction of three nuclei. (a) Schematic configuration of three nitrogen nuclei interacting equivalently with the exciplex. The disorder in an organic film suggests a distribution of $A_0$ values due to variations in the dihedral angles of phenyl groups.
(b) Weighting function for $A_N$ (c) Two-dimensional map of simulated spectra with HFI strengths $A_N$ varying from 0 MHz to 30 MHz. (d) Exemplary spectra with varying HFI strengths $A_N$. (e) By summation of all spectra from Figure (c) with weighted intensity for $A_N$, a mean value of $A_N$ = 2 MHz was found to explain the line shape of the hole spectrum.

Here, we calculate the triplet spectra for an exciplex interacting with three nuclei (Figure S7a) with a variation of $A_0$ from 0 MHz to 30 MHz and a distribution with $A_{N,max}$ = 2 MHz and $A_{N,FWHM}$ = 20 MHz. The distribution is shown in Figure S7b. Exemplary spectra are shown in Figure S7c and d. The summation of the spectra by the weighted distribution of (b) is shown in Figure S7e (black curve). Maximum values up to 30 MHz are needed to explain the broad base of the hole spectrum. If the mean value of the distribution is shifted to higher values, discrete hyperfine shoulders form in the simulation in Figure S7e. Since these features are not observed in the hole spectrum, we conclude $A_N \leq 2$ MHz.

**Spectral Broadening due to $\Delta g$-Mechanism**

An additional broadening mechanism arises, when the electronic environments of electrons on the acceptor and of holes on the donor molecules are different. A non-negligible difference in the $g$-factor is noticeable as the g-factor determines the slope of the electronic Zeeman interaction.

The influence on the ELDMR linewidth can be tested, by comparing ELDMR spectra measured at low and high resonant magnetic field positions or resonance frequencies, respectively (Figure S8a). Figure S8b shows the first derivative spectra of three exemplary ELDMR measurements at different resonance positions. A global fit is performed with EasySpin simultaneously on all three spectra to evaluate the influence of $\Delta g$. The global fit is more sensitive on the first derivative, because $\Delta g$-broadening then affects the peak positions instead of the flank width (FWHM) of the ELDMR spectrum. The global fit reveals a value of $\Delta g = 5.3 \cdot 10^{-7}$, which is a negligible influence of $\Delta g$ on the linewidth. This fact is visualized by the black arrows. The peaks show no broadening with increasing frequency since the peak position is not shifting. A higher $\Delta g$ would influence the peak position and thus ELDMR linewidth.

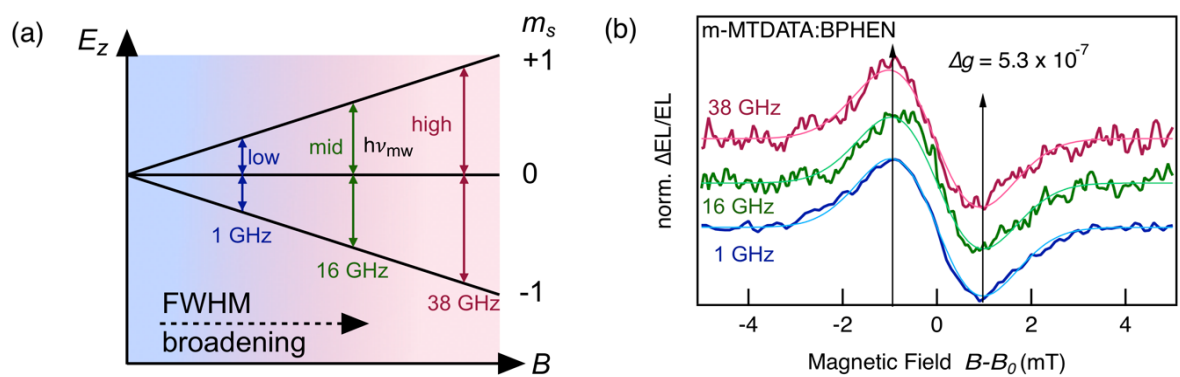

**Figure S8.** Magnetic field dependence of the ELDMR spectrum. (a) Exemplary magnetic resonance fields for low, mid and high microwave frequencies in triplet sub-levels which are split by Zeeman interaction. (b) First derivative of ELDMR spectra at 1 GHz, 16 GHz and 38 GHz. The magnetic field axis is shifted to coincide at $B - B_0 = 0$ mT to facilitate comparison. An EasySpin (thin traces) global fit is performed simultaneously on all three spectra to extract a value for $\Delta g = 5.3 \cdot 10^{-7}$. Consequently, there is no relevant change in linewidth with increasing resonance frequency and magnetic field.

## Microwave Beating by Application of Two Microwave Fields

In the hole burning experiment, two microwave fields are applied at the same time. The resulting interference is called microwave beating and the beat frequency $\Delta \nu$ is given by the difference of the two individual frequencies $\Delta \nu = |\nu_1 - \nu_2|$. A direct time-resolved measurement of the microwave beating can be realized with an oscilloscope. Due to technical limitations in time resolution, we exemplarily chose frequencies in the MHz range, to visualize the beating. The two microwave sources are merged by a microwave combiner and connected directly an oscilloscope. The beating is shown for three different beat frequencies in Figure S9.

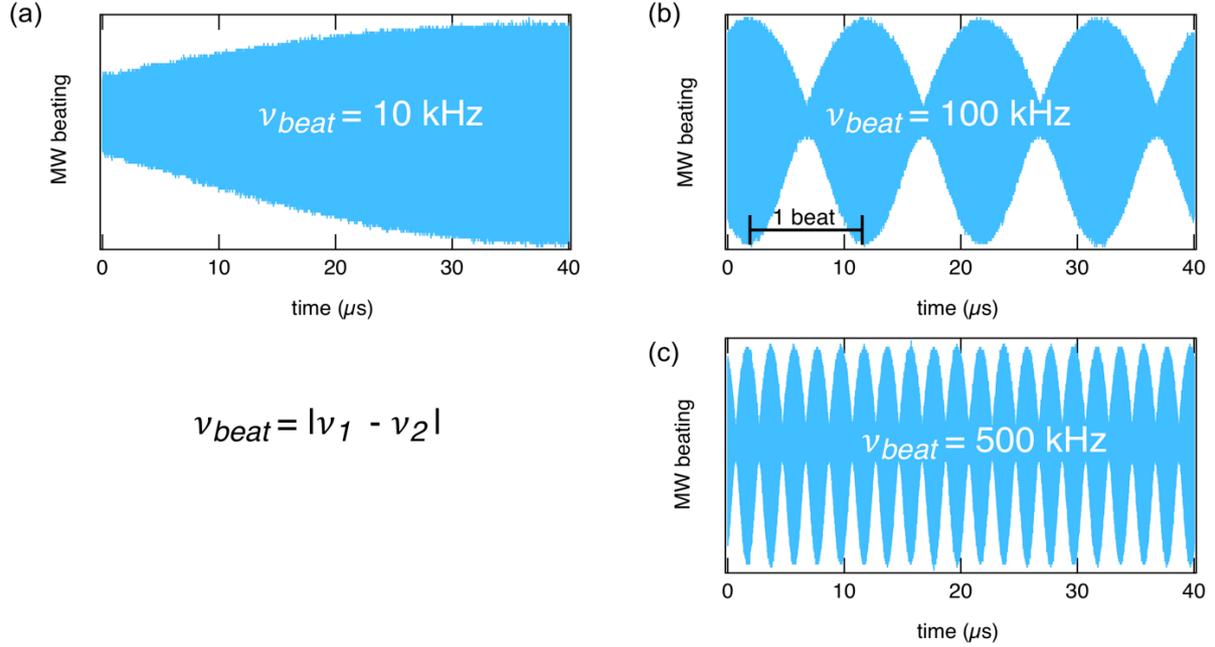

**Figure S9.** Exemplary measurement of microwave beating. The beating frequency is given by $\Delta \nu = |\nu_1 - \nu_2|$. To visualize microwave beating with an oscilloscope, frequencies in the MHz range were chosen as examples. $\nu_1$=100.00 MHz and (a) $\nu_2$=100.01 MHz (b) $\nu_2$=100.10 MHz (c) $\nu_2$=100.50 MHz.

## Simulation of Coherent Population Oscillation (CPO)

The amplitude modulation results in a brightness modulation of the OLED if the beat frequency hits critical spin relaxation time constants, such as $\nu_{beat}$ = 1/$T_1$ and 1/$T_2$. In general, the spin resonance hole burning experiment in some way samples the temporal behavior of the spin system with the beat frequency $\Delta \nu$.

To simulate this phenomenon in the hole burning spectrum, Mrozek et al. (*27*) applied a rate equation model of a two-level quantum system, which oscillates at the beat frequency $\nu_{beat}$ between $\nu_{pump}$ and $\nu_{sweep}$. In our case the quantum system is more complex and for the exciplex triplet state we use a three-level system with a small zero-field splitting. Accordingly, we adapted and expanded the model of Mrozek et al. to such a three-level system. Let us denote the triplet population $n_i^0$ of the i-th triplet state in a quasi-equilibrium state under continuous microwave radiation $p(t) = p_0$, without externally driven oscillation. Whereas the time-dependent populations of the triplet states under an oscillating microwave field $p(t) \sim \cos(2\pi \nu_{beat})$ are given as $\rho i$. The change of the triplet population in time is given by $\dot{\rho}_i$. Additionally, the relaxation rates of these levels to recover equilibrium conditions are considered as $\gamma_i$. An oscillating microwave field $p(t)$ drives transitions between the triplet states at the beat frequency $\nu_{beat}$. In the experiment, the net direction of a transition (absorption vs. emission) between the sub-levels depends on the sign of the population difference of the sub-levels. Thus, we can define the rate equation of each triplet sub-level as

$$\dot{\rho}_+ = -\gamma_+ (\rho_+ - n_+^0) - p(t)(\rho_+ - \rho_0) \qquad (S2.1)$$
$$\dot{\rho}_0 = -\gamma_0 (\rho_0 - n_0^0) + p(t)(\rho_+ - \rho_0) - p(t)(\rho_0 - \rho_-) \qquad (S2.2)$$
$$\dot{\rho}_- = -\gamma_- (\rho_- - n_-^0) + p(t)(\rho_0 - \rho_-) \qquad (S2.3)$$

Where the time dependent microwave field $p(t)$ is given by (*27*):

$$p(t) = \left\{1 + F \frac{\gamma^2}{\gamma^2 + \Delta^2} \cos[(\omega_0 + \Delta)t + \phi]\right\} p_0 \qquad (S3)$$

Where $p_0$ is the pump rate of the system and $\Delta$ is the detuning between the pump and the probe frequencies. $F$ is a contrast ratio between the pump and the probe powers. The value $\gamma$ is calculated by $\frac{1}{\gamma} = \frac{1}{\gamma_+} + \frac{1}{\gamma_0} + \frac{1}{\gamma_-}$. Figure S10 shows a simulation of Equations S2 with the relaxation rates $\gamma_+$ = 2 MHz, $\gamma_0$ = 10 kHz, $\gamma_-$ = 220 kHz, $p_0$ = 222 Hz, $F$ = 2 For high values of $\Delta$, the second term in Equation S3 vanishes and $p(t)$ retains a constant value, while the triplet system approaches equilibrium. Figure S10 (g) and (h) show the population differences given in Equations S2.

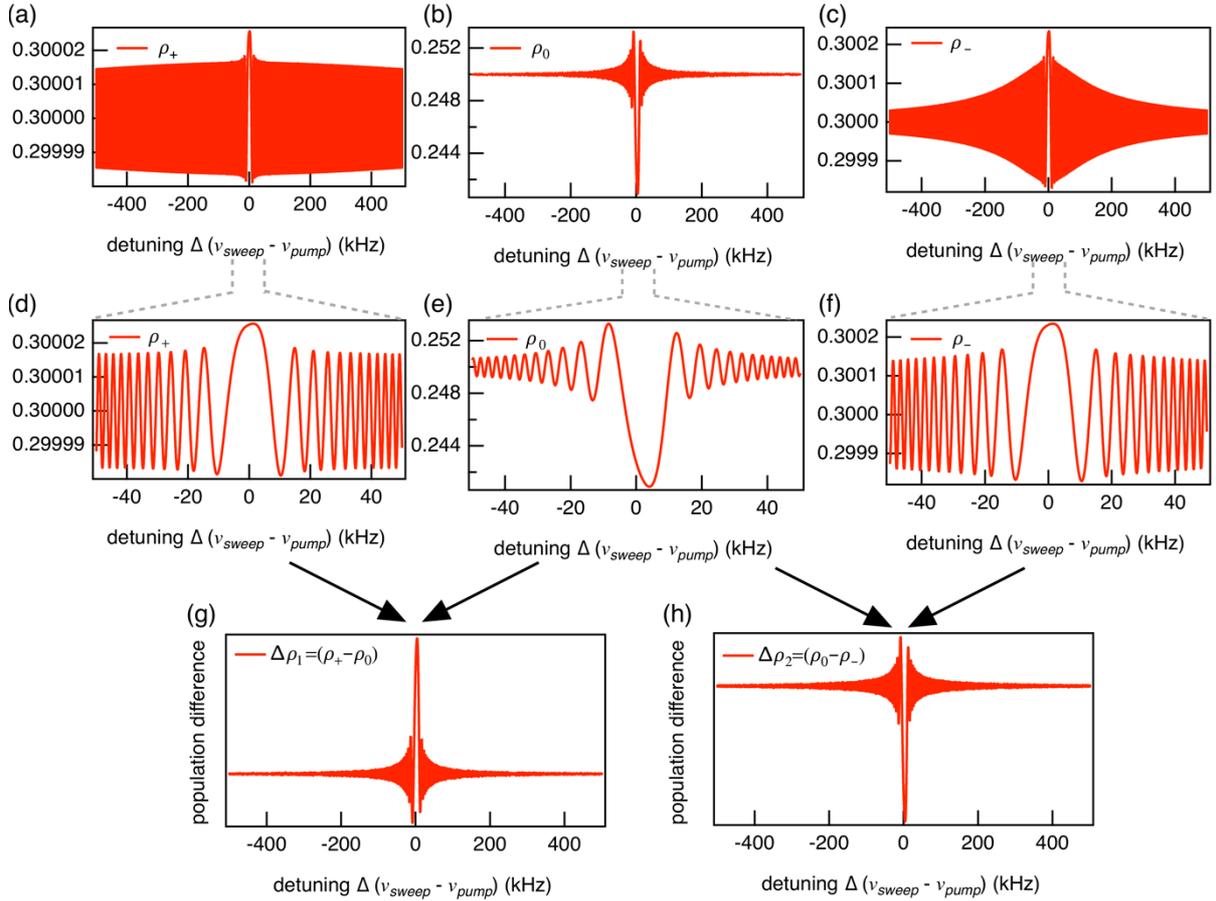

**Figure S10.** Simulation of the Coherent Population Oscillations from Equation S1. (a)-(c) Simulation of the triplet populations. (d)-(f) zoomed-in populations. (g) and (h) population difference $\Delta\rho_1 = (\rho_+ - \rho_0)$ between the $m_s = 0$ and +1 states and the difference $\Delta\rho_2 = (\rho_0 - \rho_-)$ between the $m_s = 0$ and -1 states. The envelope depends on the spin relaxation rates of the triplet sub-levels.

Since in the present study the hole burning signal was detected via lock-in amplification, the oscillations cannot be detected directly. The lock-in averages over several phases $\phi$ of the oscillation. A direct observation of the oscillation with an oscilloscope like done by Mrozek et al. was not achievable due to insufficient signal to noise ratio. To mimic the lock-in detection, we calculate the absolute value of the sum of the simulated population differences as CPO Intensity = $|\Delta\rho_1 + \Delta\rho_2|$ and applied a low-pass filter.

Multiple frequency contributions to the relaxation process are not surprising due to inhomogeneously broadened triplet states. In the simulations performed above, we summarized the three contributions by the relaxations rates $\gamma_-, \gamma_0$ and $\gamma_+$. The slowest contribution is resolved in Figure 4c from which we determine the lower limit of the spin-lattice relaxation time $T_1$. Figure S11 shows the CPO Intensity for three contributions to $T_1$ together with the ELDMR measurement.

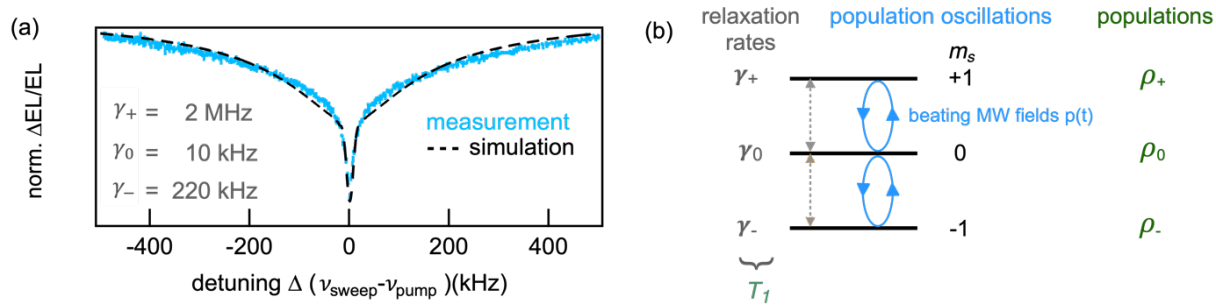

**Figure S11.** Comparison of hole burning spectra and simulation. (a) The low-pass filtered data (dashed line) of simulations in Figure S10g and h matches the lock-in detected ELDMR signal. (b) The beating MW fields result in an oscillation of the microwave power. The time-dependent power $p(t)$ creates a forced oscillation of the triplet populations. The oscillation frequency of each sub-level is limited by the relaxation rate of each level.

**Temperature-Dependence of $T_1$**

In our setup, we need to use unencapsulated OLEDs due to space constraints. Lowering the temperature makes the device stable for the week-long measurements and the acquired data set consistent. Therefore, we performed most of the measurements at a temperature of $T$ = 220 K. Additional temperature-dependent measurements for the $T_1$ time are shown in Figure S12. The dependence turns out to be weak with an inverse square-root dependence of the spin-lattice relaxation rate vs. temperature: $T_1^{-1} \sim T^{-1/2}$.

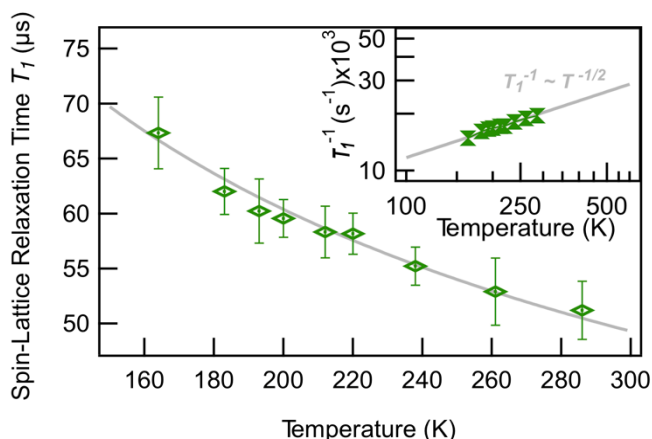

**Figure S12:** Spin-lattice relaxation time $T_1$ in dependance of experimental temperature $T$. A clear increase in relaxation time is observed for decreasing temperatures. The inset shows a log-log plot for the relaxation rate $T_1^{-1}$ with an empiric inverse square-root trend: $T_1^{-1} \sim T^{-1/2}$.

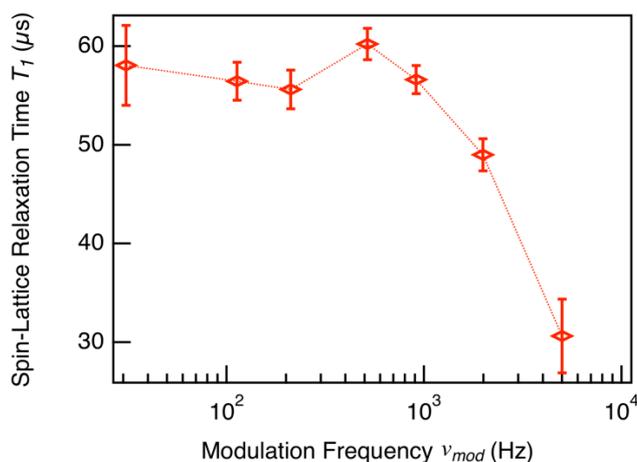

**Figure S13:** Spin-lattice relaxation time $T_1$ in dependance of the microwave modulation frequency $\nu_{mod}$. A clear decrease in relaxation time is observed if modulation of the microwaves is too fast. For all measurements we thus used $\nu_{mod}$ = 911 Hz to void this experimental artefact.